# Chaotic hypothesis: Onsager reciprocity and fluctuation–dissipation theorem*

*Giovanni Gallavotti*
Fisica, Università di Roma La Sapienza, P.le Moro 2, 00185, Roma, Italia.

*Abstract: It is shown that the "chaoticity hypothesis", analogous to Ruelle's principle for turbulence and recently introduced in statistical mechanics, implies the Onsager reciprocity and the fluctuation dissipation theorem in various reversible models for coexisting transport phenomena.*

§1: *Introduction.*

In reference [GC2] we introduced, as a principle holding when a system motions have an empirically chaotic nature, that:

*Chaotic hypothesis: A many particle system in a stationary state can be regarded as a smooth dynamical system with a transitive axiom A global attractor for the purpose of computing macroscopic properties. In the reversible case it can be regarded, for the same purposes, as a smooth transitive Anosov system.*

See [AA],[S],[R1],[Bo] for the notion and properties of Anosov systems and [Bo,R2] for the more general systems with Axiom A attractors. I do not know examples of reversible systems with an Axiom A transitive global attractor which are not transitive Anosov systems: in this sense I regard the second part of the hypothesis as a likely consequence of the first (the viceversa is trivial). In this paper only the latter part of the hypothesis will be needed and used.

The flow $x \to V_t x$ solving in phase space the differential equations of motion generates an evolution $t \to F(V_t x)$ on the observables $F(x)$. It can be naturally interpreted as a random process when the initial data $x$ are chosen randomly with a given distribution $\mu_0$.

The averages over the time variable $t$ give the *stationary* state for the evolution $V_t$, *provided they are uniquely defined*, i.e. provided for almost all choices of $x$ with distribution $\mu_0$ the averages exist and are $x$–independent. In this case the stationary state is a stationary probability distribution $\mu$: to stress that it is *dependent on $\mu_0$* we call it the *statistics $\mu$ of $\mu_0$*.

In many applications it is more convenient to regard the evolution as a discrete trasformation defined on a restricted phase space $\mathcal{C}$ of *observed events*, also called *timing events*, (which could be, for instance, the occurrence of a microscopic binary "collision"). The time evolution, or the *dynamics*, is a map $S$ of $\mathcal{C}$ into itself. The map $S$ is derived from the flow $V_t$ solving the differential equations of motion of the system: the timing events $\mathcal{C}$ have to be thought as a surface transversal to the flow and if $t(x)$ is the time between the timing event $x$ and the

---

successive one $Sx$ it is $V_{t(x)}x = Sx$. Note that the points $V_t x$ *are not* timing events (*i.e.* they are not in $\mathcal{C}$) for the intermediate times $0 < t < t(x)$.

The notions of *statistics* $\mu$ of $\mu_0$ carry over unchanged to the discrete notions of phase space and of evolution and the above chaotic hypothesis is assumed in such a context. When referring to phase space, unless stated otherwise, we think of a phase space $\mathcal{C}$ consisting of timing events and of a map $S$ on $\mathcal{C}$ defining the time evolution. The *smoothness* of the Anosov system is intended in all the coordinates and parameters on which the system equations depend.

The chaotic hypothesis implies, as a mathematical consequence, that *for most random distributions $\mu_0$ for the choice of the initial data $x$ the distribution $\mu$ exists*.

However the choice of the initial data with distribution $\mu_0$ *proportional to the volume measure on $\mathcal{C}$* plays a special role, because in the case of hamiltonian systems such distribution is generated naturally via the microcanonical ensemble.

For instance one can read, translating symbols into the present notations: "the appropriate objects of study of a statistical mechanics of time dependent phenomena are the random processes $F(V_t x)$ *with initial distribution of $x$, $\rho_E(x)$ (the microcanonical distribution)*, for all energies of interest and for all gross variables $F$ of interest", (italics added), see the nice paper [GCDR].

I "strongly disagree" with such a philosophical position: but I will adopt it here because it is obvious that, whether one agrees or not, it is *of fundamental interest* to understand the statistics of initial data chosen at random with a distribution proportional to the phase space volume. Other random choices may have to wait until the latter is properly understood.

In the Physics literature the existence of the distribution $\mu$ is, in fact, assumed in general as stated by the following (extension) of the zeroth law, [UF], giving a global property of the motions generated by initial data chosen randomly with distribution $\mu_0$ proportional to the volume measure on $\mathcal{C}$:

*Extended zero-th law: A dynamical system $(\mathcal{C}, S)$ describing a many particle system (or a continuum such as a fluid) generates motions that admit a statistics $\mu$ in the sense that, given any (smooth) macroscopic observable $F$ defined on the points $x$ of the phase space $\mathcal{C}$, the time average of $F$ exists for all $\mu_0$-randomly-chosen initial data $x$ and is given by:*

$$\lim_{M \to \infty} \frac{1}{M} \sum_{k=0}^{M-1} F(S^j x) = \int_{\mathcal{C}} \mu(dx') F(x') \tag{1.1}$$

*where $\mu$ is a $S$-invariant probability distribution on $\mathcal{C}$.*

The chaotic hypothesis was proposed by Ruelle in the case of fluid turbulence, and it is extended to non equilibrium many particle systems in [GC1]. If one assumes it, then it *follows* that the zeroth law holds, [S,Bo,R]; however it is convenient to regard the two statements as distinct because the hypothesis we make is "*only*" that one can suppose that the system is Axiom A or Anosov for "practical purposes": this leaves the possibility that it is not strictly speaking such and some ("negligible in the thermodynamic limit") corrections may be needed on the predictions obtained by using the hypothesis.

From now on only reversible systems will be considered in this paper: they are dynamical systems such that there is an isometric map $i$ of phase space such that $i^2 = 1$ and $iS = S^{-1}i$. In [GC2] the generality of the hypothesis is discussed and in [GC1], [GC2] we derived, as a rather general consequence, predictions testable by numerical experiment in systems with few degrees of freedom. The most relevant feature of the prediction, which is a *large deviation theorem* or *fluctuation theorem* for systems with reversible dynamics, is that it is *parameter free*.

A drawback is that it *is not* testable directly in really large systems.

The question of whether the "chaotic hypothesis" could be tested on real experiments, *i.e.* for really macroscopic systems, remained open. In this paper we show, through examples, that the chaotic hypothesis implies quite generally, in systems with reversible dynamics, the Onsager reciprocity and the fluctuation–dissipation theorem: see [DGM] for a classical general discussion of the reciprocity relations, see [C] (Appendix A, p.187-200) for a kinetic derivation, see also [E,ELS, GJL] for recent developments. A very nice introduction and an exposition of the basic ideas can be found in [D]. The ideas of the present paper can be applied also to systems relevant for the theory of developed turbulence (see [G4]).

A puzzling aspect of the chaotic hypothesis is that it implies that the system has a positive gap separating from 0 the Lyapunov exponents and one may have serious doubts on the validity of such a strong property: so that a discussion is in order.

This is discussed in [GC2], §8, and [G4] by suggesting that there may well be many vanishing exponents (or exponents of order $O(N^{-1})$) if $N$ is the particle number: such exponents should be "ignored" as they should correspond to macroscopic evolution laws which microscopically will be effectively described by local conservation laws [EMY,KV,LOY]. They have an approximate character, unless $N \to \infty$, but one can think of imposing them as *exact* conservation rules by adding to the forces acting on the system other suitable "auxiliary" forces *minimally* required to achieve the purpose of turning the slow macroscopic observables responsible for the existence of the 0 Lyapunov exponents into exact local conservation rules. For instance one can find the auxiliary forces by applying the Gauss' *principle of least constraint*, see appendix. The dynamical system obtained in this way should be one to which the chaotic hypothesis should apply.

In §2,§3 we introduce some models which will be used to illustrate our general ideas; in §4 we discuss the relevant mathematical facts about reversible Anosov systems; in §5 we give a proof, whose full mathematical rigor still rests on a mathematical conjecture (§5) on Anosov systems (that I hope to address elsewhere), of the validity of the fluctuation dissipation relation and of Onsager reciprocity in the models introduced in §2,§3: but the generality of the argument, which seems largely model independent, will also emerge. Comments and a critical comparison with the classical derivations (that apply to our models as well) are presented in §6.

*§2: A diffusion problem.*

We consider a mixture of two chemically inert gases whose $N = N_1 + N_2$ molecules (with equal masses $m$ and respective numbers $N_1 = N_2$) are contained in a box $\mathcal{B} = [-\frac{1}{2}L, \frac{1}{2}L]^2$ with periodic boundary conditions and are subject to a respective force field $\underline{E}^1 = E^1 \underline{i}$ and $\underline{E}^2 = E^2 \underline{i}$. The molecules interact via a pair interaction with a short range potential (*e.g.* a Lennard Jones type of potential).

Furthermore the motion is subject to the constraint that the total energy is constant, via a constraint force law which is *ideal* in the sense that it verifies Gauss' principle of minimal constraint. This means that, if $E_j$ is the field on the $j$-th particle (equal either to $E^1$ or to $E^2$) the equations of motion are:

$$\underline{\dot{q}}_j = \frac{1}{m} \underline{p}_j, \qquad \underline{\dot{p}}_j = \underline{F}_j + E_j \underline{i} - \alpha \underline{p}_j \tag{2.1}$$

with $\alpha = \dfrac{\sum_j E_j \underline{i} \cdot \underline{p}_j}{\sum_j \underline{p}_j^{\,2}}$, and we expect that the future time average of the total momentum $\underline{P}$ and of the dissipation $\alpha$ will be some $\langle \underline{P} \rangle_+ > 0$ and $\langle \alpha \rangle_+ > 0$ if $(E^1, E^2) \neq 0$.

The average is expected to be attained exponentially fast while the "conjugate" variable $\underline{C}$, the center of mass position, will be expected to evolve with zero Lyapunov exponent (and to behave asymptotically as $\langle \underline{C} \rangle_+ = \frac{1}{(N_1+N_2)m} \langle \underline{P} \rangle_+ t + cost)$.

According to the analysis of [GC2],§8, and [G4], we expect that we can freely add to the equations of motion a minimal constraint force imposing the constraint $\underline{P} = P(E^1, E^2) \underline{i}$ if $P(E^1, E^2)$ is the average horizontal momentum, *i.e.* the (unknown) "equation of state", (and the implied $\underline{\dot C} = cost$). This means that considering a modified equation of motion:

$$\underline{\dot q}_j = \frac{1}{m}\underline{p}_j, \qquad \underline{\dot p}_j = \underline{F}_j + E_j\,\underline{i} - \alpha\,\underline{p}_j - \underline{\beta} \qquad (2.2)$$

with $\underline{\beta} = \frac{E_1 N_1 + E_2 N_2}{N_1 + N_2}\,\underline{i} - \alpha\frac{\underline{P}}{N_1+N_2}$, should not lead to appreciably different qualitative behaviour if the initial data are consistent with the equation of state $\underline{P} = P(E^1, E^2)\,\underline{i}$, $\underline{C} = \underline{0}$.

The phase space contraction per unit time is:

$$\gamma = (2(N_1 + N_2)(1 - \frac{1}{N_1+N_2}) - 1)\,\alpha = (2(N_1+N_2) - 3)\,\alpha \qquad (2.3)$$

Note that without imposing the extra constraint the phase spave contraction would have been $(2(N_1+N_2) - 1)\,\alpha$: with a relative difference of $O(N^{-1})$, which should become neglegible in the thermodynamic limit $L \to \infty, N_1 L^{-2} = \rho_1, N_2 L^{-2} = \rho_2$.

We call $Q$ the work per unit time performed by the forces $\alpha\underline{p}_j$ so that the phase space contraction rate in the configuration $x$ is (to leading order in $N$) $2N\alpha(x)$, *i.e.* $\frac{E^1 N_1 \dot a_1(x) + E^2 N_2 \dot a_2(x)}{k_B T(x)}$ = $\frac{Q}{k_B T}$ if $k_B$ is Boltzmann's constant, $k_B T(x)$ is the kinetic energy per particle and $a_1(x), a_2(x)$ are the horizontal coordinates of the centers of mass $\underline{C}_1, \underline{C}_2$ of the two species, and $\dot a_1(x), \dot a_2(x)$ the corresponding velocities. Thus the phase space contraction rate can be interpreted as $\frac{1}{k_B}$ times the *entropy creation rate* (at least if $E^1, E^2$ are so small that the centers of mass horizontal velocities are small compared to the root mean square velocities).

The above model is closely related to the *color diffusion* model considered in [BEC]. A *very important* feature of the model is its *time reversibility*: the map $i(\underline{p}, \underline{q}) = (-\underline{p}, \underline{q})$ has the property that it is an isometry of phase space such that $iS = S^{-1}i$ and $i^2 = 1$.

§3: *A heat conduction-electrical conduction model.*

As a second model we consider a modification of model 4 of [CG2], inspired by [HHP], see also [PH] for a more general perspective on constrained systems. This is a model for a heat conducting and electrically conducting gas. In a box $\mathcal{B} = [-2L - H, 2L + H] \times [-\frac{1}{2}L, \frac{1}{2}L]$ are enclosed $N$ particles with mass $m$, interacting via a rather general pair potential, like a hard core potential with a tail or via a Lennard Jones potential, and they are subject to a constant force field (*electric field*) $E\,\underline{i}$ in the $x$ direction. The boundary conditions are periodic in the horizontal direction and reflecting in the vertical direction.

Adjacent to the box $\mathcal{B}$ there are two boxes $\mathcal{R}_+, \mathcal{R}_-$ containing $N_+ = N_-$ particles interacting with each other via a hard core interaction, and with the particles in $\mathcal{B}$ via a pair interaction (*e.g.* with potential equal to the one between the particles in $\mathcal{B}$), but are separated from the latter by a reflecting wall.

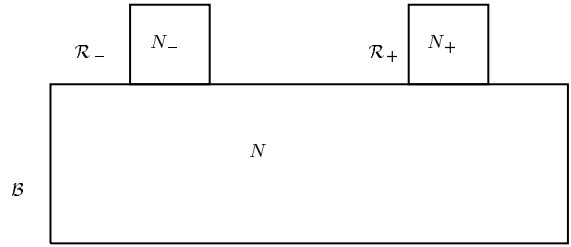

The model name is motivated because we imagine other forces to act on the system: they are the minimal forces (in the sense of Gauss' minimal constraint principle, see appendix A1) necessary to enforce the following constraints:

1) the total kinetic energy in the "hot plate" $\mathcal{R}_+$ and in the "cold plate" $\mathcal{R}_-$ are constrained to be $N_+ k_B T_+$ and, respectively, $N_- k_B T_-$ where $T_-$ and $T_+ = T_- + \delta T, \delta T \geq 0$ are the *temperatures of the plates*.

2) the total energy $U$ in the box $\mathcal{B}$ is constrained to stay constant.

The equations of motion are:

$$\begin{aligned}\dot{\underline{q}}_j &= \frac{\underline{p}_j}{m} \\ \dot{\underline{p}}_j &= \underline{F}_j + E\chi(\underline{q}_j)\underline{i} - \alpha_+ \chi_+(\underline{q}_j)\underline{p}_j - \alpha_- \chi_-(\underline{q}_j)\underline{p}_j - \alpha\chi(\underline{q}_j)\underline{p}_j\end{aligned} \quad (3.1)$$

where $\chi, \chi_\pm$ are the characteristic functions of the regions $\mathcal{B}, \mathcal{R}_\pm$ and $\alpha_+, \alpha_-, \alpha$ are multipliers defined so that for some $T_\pm, U$:

$$\sum_{j=1}^{N_\pm} \chi_\pm(\underline{q}_j)\frac{\underline{p}_j^2}{2m} = N_\pm k_B T_\pm, \qquad \sum_{j=1}^{N} \chi_\mathcal{B}(\underline{q}_j)\frac{\underline{p}_j^2}{2m} + V(\underline{q}) = U \quad (3.2)$$

are exact constants of motion.

We suppose for simplicity (see however comment (1) in §6 below) that the system in $\mathcal{B}$ is kept at a constant total energy $U$, and at constant reservoirs temperatures $T_\pm$. In this case we call $Q_+, Q_-, Q_0$ the work per unit time performed by the forces $\alpha_+\chi_+\underline{p}_j, \alpha_-\chi_-\underline{p}_j, \alpha\chi\underline{p}_j$ respectively.

Let $\mathcal{L}_E, \mathcal{L}_+^0, \mathcal{L}_-^0, \mathcal{L}_+, \mathcal{L}_-$ denote, respectively, the work per unit time performed by the field $E$ or by the particles in the thermostats $\mathcal{R}_+, \mathcal{R}_-$ on the gas in $\mathcal{B}$, or by the gas in $\mathcal{B}$ on the thermostats $\mathcal{R}_+, \mathcal{R}_-$. Then the imposed conservation laws give:

$$\begin{aligned}-Q_0 + \mathcal{L}_E + \mathcal{L}_+ + \mathcal{L}_- &= 0 &&\longleftrightarrow&& (U = const) \\ -Q_+ + \mathcal{L}_+^0 &= 0 &&\longleftrightarrow&& \Big(\sum_{\underline{q}_j \in \mathcal{R}_+} \frac{\underline{p}_j^2}{2m} = N_+ k_B T_+\Big) \\ -Q_+ - \mathcal{L}_-^0 &= 0 &&\longleftrightarrow&& \Big(\sum_{\underline{q}_j \in \mathcal{R}_-} \frac{\underline{p}_j^2}{2m} = N_- k_B T_-\Big)\end{aligned} \quad (3.3)$$

so that one easily finds (by differentiating (3.2) with respect to time and by applying the equations of motion, (3.1)) expressions for $\alpha_+, \alpha_-, \alpha$:

$$\alpha = \frac{\mathcal{L}_E + \mathcal{L}_+ + \mathcal{L}_-}{\sum_{\underline{q}_j \in \mathcal{B}} \underline{p}_j^2/m} = \frac{Q_0}{\sum_{\underline{q}_j \in \mathcal{B}} \underline{p}_j^2/m}, \qquad \alpha_\pm = \frac{\mathcal{L}_\pm^0}{2N_\pm k_B T_\pm} = \frac{Q_\pm}{2N_\pm k_B T_\pm} \quad (3.4)$$

and the phase space contraction per unit time, or $\frac{1}{k_B}$ times the *entropy creation per unit time*, is:

$$\gamma(x) = (2N_\pm - 1)(\alpha_+ + \alpha_-) + (2N - 1)\alpha \quad (3.5)$$

in the configuration $x$.

The above model also shares the feature that it is *time reversible*, and the isometric map $i(\underline{p}, \underline{q}) = (-\underline{p}, \underline{q})$ is again such that $iS = S^{-1}i$, and $i^2 = 1$.

### §4: The SRB distribution.

The chaotic hypothesis of §2 allows us to represent the SRB distribution in a simple form, by using *Markov partitions*, [S]. We consider *only* transitive reversible Anosov systems, although many concepts make sense for more general systems with chaotic attractors.

The notion of Markov partition is a mathematically precise version of the intuitive idea of *coarse graining*. We just recall here the main properties of Markov partitions. For a discussion of the intuitive meaning and the connection with the coarse graining see [G3].

1) A parallelogram will be a small set with a boundary consisting of pieces of the stable and unstable manifolds of the map $S$ joined together as described below. The smallness has to be such that the parts of the manifolds involved look essentially "flat": *i.e.* the sizes of the sides have to be small compared to the smallest radii of curvature of the *unstable manifolds* $W_x^u$ or of the *stable manifolds* $W_x^s$, as $x$ varies in $\mathcal{C}$.

Therefore let $\delta$ be a length scale small compared to the minimal (among all $x$) curvature radii of the stable and unstable manifolds. Let $\Delta^u$ and $\Delta^s$ be small (and "small" means of size $\ll \delta$) connected surface elements on $W_x^u$ and $W_x^s$ containing $x$. We define a *parallelogram $E$* in the phase space $\mathcal{C}$, to be denoted by $\Delta^u \times \Delta^s$, with center $x$ and axes $\Delta^u$, $\Delta^s$ as follows.

Consider $\xi \in \Delta^u$ and $\eta \in \Delta^s$ and suppose that the point $z$, denoted $\xi \times \eta$, such that the shortest path joining $\xi$ with $\eta$ formed by a path on the stable manifold $W_\xi^s$ joining $\xi$ to $z$ and by a path on the unstable manifold $W_\eta^u$ joining $z$ to $\eta$, is uniquely defined. This will be so if $\delta$ is small enough and if $\Delta^u$, $\Delta^s$ are small enough compared to $\delta$ as we assume, (because the stable and unstable manifolds are "smooth" and transversal).

The set $E = \Delta^u \times \Delta^s$ of all the points generated in this way when $\xi, \eta$ vary arbitrarily in $\Delta^u, \Delta^s$ is called a parallelogram (or rectangle), provided the boundaries $\partial \Delta^u, \partial \Delta^s$ of $\Delta^u$ and $\Delta^s$ as subsets of $W_x^u$ and $W_x^s$, respectively, have zero surface area on the manifolds on which they lie. The sets $\partial_u E \equiv \Delta^u \times \partial \Delta^s$ and $\partial_s E = \partial \Delta^u \times \Delta^s$ will be called the *unstable* or *horizontal* and *stable* or *vertical* sides of the parallelogram $E$.

Consider a partition $\mathcal{E} = (E_1, \ldots, E_\mathcal{N})$ of $\mathcal{C}$ into $\mathcal{N}$ rectangles $E_j$ with pairwise disjoint interiors. We call $\partial_u \mathcal{E} \equiv \cup_j \partial_u E_j$ and $\partial_s \mathcal{E} \equiv \cup_j \partial_s E_j$: these are called respectively the *unstable boundary* of $\mathcal{E}$ and the *stable boundary* of $\mathcal{E}$, or also the horizontal and vertical boundaries of $\mathcal{E}$, respectively.

2) We say that $\mathcal{E}$ is a *Markov partition* if the transformation $S$ acting on the stable boundary of $\mathcal{E}$ maps it into itself (this means $S\partial_s \mathcal{E} \subset \partial_s \mathcal{E}$) and if, likewise, the map $S^{-1}$ acting on the unstable boundary maps it into itself ($S^{-1} \partial_u \mathcal{E} \subset \partial_u \mathcal{E}$).

The actual construction of the SRB distribution then proceeds from the important result of the theory of Anosov systems expressed by a remarkable theorem (Sinai, [S]):

*Theorem: Every Anosov system admits a Markov partition $\mathcal{E}$.*

*Comments:*
a) If the reversibility property holds it is clear that $i\mathcal{E}$ is also a a Markov partition. This follows from the definition of Markov partition and from the fact that reversibility implies:

$$W_x^s = iW_{ix}^u \qquad (4.1)$$

b) The definition of a Markov partition also implies that the intersection of two Markov partitions is a Markov partition, hence it is clear that if a transitive Anosov system is reversible (*i.e.* there is an isometry $i$ such that $iS = S^{-1}i$ and $i^2 = 1$) there are Markov partitions $\mathcal{E}$ that are reversible in the sense that $\mathcal{E} = i\mathcal{E}$, *i.e.* if $E_j \in \mathcal{E}$ there is $j'$ such that $iE_j = E_{j'} \in \mathcal{E}$.

c) The usefulness of the Markov partitions comes from the possibility that they provide of representing the points of $\mathcal{C}$ as *infinite strings* of symbols (in a more useful way than representing them, for instance, as the strings of digits that give the value of their coordinates).

This is simply achieved by associating with $x \in \mathcal{C}$ the string $\underline{j} = \{j_k\}_{k=-\infty}^{\infty}$, such that $S^k x \in E_{j_k}$. The invertibility of the map between $x \in \mathcal{C}$ and the *compatible* or *allowed* sequences, *i.e.* the sequences $\underline{j}$ such that the interior of $SE_{j_k}$ intersects the interior of $E_{j_{k+1}}$ is a well known (and easy) consequence of the definition of Markov partition. The correspondence is in fact one to one with some obvious exceptions: namely to each sequence $\underline{j}$ with the above compatibility property corresponds one $x$; viceversa, if $x$ is not on a boundary of some of the $E \in \mathcal{E}$ nor on the image of a boundary under a power of $S$ then $x$ admits only one symbolic representation. The points on the boundaries or visiting, in their evolution, the boundaries of course have several (but finitely many) symbolic representations. Just in the same way as the decimal representation of a number is unique for most numbers: the ones which end with infinitely many successive 9 admit two representations.

The correspondence $\underline{j} \longleftrightarrow x$ between points $x \in \mathcal{C}$ and their history, or symbolic representation, $\underline{j}$ as a compatible sequence will be denoted $x = x(\underline{j})$ (*symbolic code*).

d) If we define the *compatibily matrix*, or *intersection matrix*, $C_{ij}$ by setting $C_{ij} = 1$ if the interior of $E_j$ intersects the interior of $SE_i$ and $C_{ij} = 0$ otherwise, then the assumed transitivity implies that there is a iterate $q$ of $C$ such that all elements of $C^q$ are positive (*i.e.* $S^q E_j$ has interior intersecting $E_k$ for all pairs $j, k$, simoultaneosly).

e) Consider the partition $\mathcal{E}_M = \cap_{-M}^{M} S^{-j} \mathcal{E}$ obtained by intersecting the images under $S^k$, $k = -M, \ldots, M$, of $\mathcal{E}$. Then $\mathcal{E}_M$ is still a Markov partition and it is time reversal invariant if $\mathcal{E}$ is (see (b) above). Note that the parallelograms of $\mathcal{E}_M$ can be labeled by the strings of symbols $j_{-M}, \ldots, j_M$ and they consist of the points $x$ such that $S^k x \in E_{j_k}$ for $-M \leq k \leq M$. In other words the parallelograms consist of those points $x$ which, in their time evolution, visit at time $k$ the parallelogram $j_k$.

f) If $F(x)$ is a function on phase space (*observable*) then we can regard it as a function $F(x(\underline{j}))$ on symbolic sequences. An observable $F$ is *local* if it "depends exponentially little" from the history symbols $j_k$ with large $k$: *i.e.* if $F(x(\underline{j})) - F(x(\underline{j}\,'))$ tends to zero exponentially fast with the maximum number $k$ such that $\underline{j}$ and $\underline{j}\,'$ agree on the sites with label $h$ with $|h| \leq k$. A simple condition on $F$ guaranteeing its locality is that $F$ is Hölder continuous in $x$.

We now construct a probability distribution on $\mathcal{C}$ by defining it as a probability distribution on the space of the compatible strings $\underline{j}$ and then by interpreting it as a distribution on the phase space $\mathcal{C}$.

(1) For this purpose we first pick a point, that we call the *center*, $x_{j_{-M}, \ldots, j_M}$ in each $E_{j_{-M}, \ldots, j_M}$ with non empty interior simply by considering the compatible string which is obtained by continuing the string $j_{-M}, \ldots, j_M$ "to the right" into a string $j_M, j_{M+1}, \ldots$ and to the "left" into a string $\ldots, j_{-M-1}, j_{-M}$ in a way that the whole string $\underline{j}$ is compatible (*i.e.* such that there is a point $x$ such that $S^k x \in E_{j_k}$ for all $k$) and, *furthermore*, the entries of the continuation strings *depend only* on the value of $j_M$ and $j_{-M}$ respectively.

In general given $j_{\pm M}$ there will be many choices of the continuation strings: which one we actually take is irrelevant. We impose however the further constraint that the continuation is made in a "time reversible way", *i.e.* we choose the continuations so that if $x$ is the center of $E_j$ then $ix$ is the center of $iE_j$. A further restriction (not necessay in the following but very nattural) that one could consider is imposing that the continuation string to the right of $j_M$ (or to the left of $j_{-M}$) agree identically after finitely many steps. Note that the existence of the continuation strings and the possibility of imposing on them the above restrictions is an immediate consequence of the transitivity property of the compatibility matrix $C$. (2) We then define, given $\tau > 0$:

$$\overline{\Lambda}_{u,\tau}(x) = \prod_{j=-\tau/2}^{\tau/2-1} \Lambda_u(S^j x) \qquad (4.2)$$

where $\Lambda_u(x)$ is the *local expansion coefficient* of the surface elements of the unstable manifold at $x$, *i.e.* it is the jacobian determinant of the transformation $S$ regarded as a map of $W_x^u$ into $W_{Sx}^u$. Likewise we define $\Lambda_s(x)$ and $\overline{\Lambda}_{s,\tau}(x)$ as the corresponding quantities obtained by regarding $S$ as a map of the stable manifold $W_x^s$ to $W_{Sx}^s$.

(3) Finally we define a distribution $\mu_{M,\tau}$ on $\mathcal{C}$ by "giving" to each set $E_{j_{-M}, \ldots, j_M} \in \mathcal{E}_M$ a

probability proportional to $\overline{\Lambda}_{u,\tau}^{-1}(x_{j_{-M},\ldots,j_M})\delta_\tau^{-1}(x_{j_{-M},\ldots,j_M})$ where $\delta(x) = \sin\vartheta(x) = \delta_0(x)$ is the sine of the angle between the stable and the unstable manifolds at $x$ and $\delta_\tau(x) = \delta(S^{\tau/2}x)$.

More precisely we define the distribution $\mu_{M,\tau}$ so that the integral of a smooth function $F$ is:

$$\int_\mathcal{C} \mu_{M,\tau}(dx)F(x) \stackrel{def}{=} \frac{\sum_j \overline{\Lambda}_{u,\tau}^{-1}(x_j)\delta_\tau^{-1}(x_j)F(x_j)}{\sum_j \overline{\Lambda}_{u,\tau}^{-1}(x_j)\delta_\tau^{-1}(x_j)} \qquad (4.3)$$

where $j$ is a short hand notation for $j_{-M},\ldots,j_M$ and $x_j = x_{j_{-M},\ldots,j_M}$ is the "center" chosen above in $E_j \in \mathcal{E}_M$. No relation is assumed here between $T$ and $\tau$, although in the applications we shall (naturally) take $T = \tau/2$, as this simplifies the discussion considerably.

The distribution $\mu_{M,\tau}$ is very interesting because it is an approximation (a very good one) of the SRB distribution. In fact in [G1,CG1,CG2] the following theorem is shown to be a reformulation (convenient although trivially equivalent) of a basic theorem by Sinai:

*Theorem: If $(\mathcal{C}, S)$ is a transitive Anosov system the SRB distribution $\mu$ exists and the $\mu$ average of a local function (see (f) above) $F$ is:*

$$\int_\mathcal{C} \mu(dx)F(x) = \lim_{M\to\infty,\tau\to\infty} \int_\mathcal{C} \mu_{M,\tau}(dx)F(x) \qquad (4.4)$$

*Furthermore in (4.3) the factor $\delta_\tau^{-1}(x_j)$ could be replaced by $\delta_\tau^z(x_j)$ with $z$ any prefixed real number (e.g. $z = 0$). The limits can be interchanged.*

The original statement is that $\mu$ exists and it is a Gibbs state with potential $\log \Lambda_u^{-1}(x)$: see [Bo],[R1],[R2],[S] for a discussion of this form of the statement. In [R2] the latter statement is extended to cover the case in which $(\mathcal{C}, S)$ has a global transitive Axiom A attractor: the discussion in [G1],[G2] shows that the above theorem extends, unchanged, to such cases. The extra factor $\delta_\tau^z$ with $z = -1$ was absent in [GC1],[GC2] where $z$ was chosen equal to 0 (an admissible alternative choice).

The possibility of fixing $z$ arbitrarily, in spite of the apparently strong modification it introduces, is easily seen by examining the proof of (4.4), see [G1], [G2]. The proof is based on the interpretation of (4.3) as a probability distribution on the space of the compatible strings. In this interpretation one immediately recognizes that (4.3) corresponds to a finite volume Gibbs distribution for a suitable short range hamiltonian defined on the space of compatible strings. An extra factor $\delta_\tau^z(x_j)$ corresponds to considering the same Gibbs distribution *just with a different boundary condition*: which becomes irrelevant in the limit as $\tau \to \infty$ because one dimensional Gibbs states with short range interactions do not have phase transitions and therefore are insensitive to changes in the boundary conditions. Different choices of the center points also correspond to different choices of boundary conditions.

The choice $z = -1$ is much better than $z = 0$ because it leads to simpler formulae and arguments: we shall call (4.3) a *balanced* approximation to the SRB distribution because as we shall see it is *reminiscent* of a probability distribution verifying the detailed balance (which however is *not* verified in our models, except in 0 forcing, *i.e.* in equilibrium).

In (4.4) with $T \gg \tau/2$ the choice of the point $x_j$ in the parallelograms of $\mathcal{E}_M$ can be really arbitrary, and it does not matter that $x_j$ is really chosen as said above or just *anywhere* in $E_{j_{-M},\ldots,j_M}$ (because the variation of the weigths (4.2) is in this case negligible, provided $M - \tau/2 \to \infty$ fast enough).

The extra properties that we need are that $\mathcal{E}_M$ is reversible (see above), *i.e.* $iE_j = E_{j'} \in \mathcal{E}_M$ (for a suitable $j'$) and that, as a consequence of the reversibility (via (4.1) and the isometric nature of the time reversal map $i$ and the validity of $\gamma_\tau(x) = -\gamma_\tau(ix)$ for the underlying differential equations):

$$\overline{\Lambda}_{u,\tau}(ix) = \overline{\Lambda}_{s,\tau}^{-1}(x), \qquad \delta_0(x) = \delta_0(ix), \qquad \delta_\tau(x) = \delta_{-\tau}(ix) \qquad (4.5)$$

which are identities (see [GC2]); for the definitions of $\Lambda_s, \overline{\Lambda}_{s,\tau}$, see the lines following (4.2). Furthermore the volume measure and the expansion and contraction rates are related by:

$$\overline{\Lambda}_{u,\tau}(x)\overline{\Lambda}_{s,\tau}(x)\frac{\delta_\tau(x)}{\delta_{-\tau}(x)} \equiv e^{-\tau t_0 \overline{\sigma}_\tau(x)} \tag{4.6}$$

where $t_0$ is the average time interval between successive timing events and the phase space volume contraction for a single transformation is written $e^{-t_0\sigma(x)}$, thus *defining* $\sigma(x)$ and $\overline{\sigma}_\tau(x)$:

$$\overline{\sigma}_\tau(x) \stackrel{def}{=} \frac{1}{\tau}\sum_{r=-\tau/2}^{\tau/2} \sigma(S^r x) \tag{4.7}$$

The (4.6) is is obtained from the relation $\Lambda_u(x)\Lambda_s(x)\frac{\delta(Sx)}{\delta(x)} = e^{-t_0\sigma(x)}$ by evaluating it on the points $S^k x$, $k = -\frac{\tau}{2}, \ldots, \frac{\tau}{2}$ and multiplying the results.

If the time interval $t(x)$ between the timing event $x \in \mathcal{C}$ and the successive one is very small and if its fluctuations can be neglected toghether with those of $\gamma(x)$ see (2.3), (3.5) (within the same time interval) then one simply has $\sigma(x) = \gamma(x)$. Note that *in all cases* with any reasonable definition of timing events the time $t(x)$ will tend to zero in the thermodynamic limit (as $O(N^{-1})$), but also $\gamma$ will tend to infinity as $O(N)$.

More generally there is a simple relation between the function $\sigma(x)$ above and the function $\gamma(x)$ which describes the phase space contraction rate in the differential equations giving rise to the map $S$ (see (2.3), (3.5)); namely:

$$\sigma(x) = \frac{1}{t_0}\int_0^{t(x)} \gamma(S_t x)\, dt \tag{4.8}$$

But the use of (4.8) is quite clumsy and one can always think that the timing events are chosen, artificially, much closer than the natural $t_0 = O(1/N)$ and observed at constant time intervals so that no difference really exixts between $\gamma(x)$ and $\sigma(x)$. If necessity arises one can always use the precise relation (4.8), at the expense of some formal algebraic complications in the intermediate step of our coming deductions.

§5: **Applications to the models. Onsager reciprocity and fluctuation–dissipation theorem.**

In this section we neglect for simplicity of exposition the difference between $\sigma(x)$ and $\gamma(x)$, *i.e.* we suppose that $t(x) = t_0$ is constant and that $\gamma$ is constant on the path traveled in the time interval $t(x)$: this simplifies considerably the algebra and the reader should have no trouble checking that the proper relation (4.8) could be consistently used leading to no corrections to the final results below (because in the end we shall set $\underline{E} = \underline{0}$).

Relation (4.3) has the form of a statistical average and we shall try to use it in the "same" way as in equilibrium statistical mechanics. We shall first study here the two currents $J_h$, $h = 1, 2$, generated by the pair of fields $E_h$ in the diffusion model of §3.

The two currents are, if $\rho_h, v_h$ are the density and average velocity of the species $h$:

$$J_h = \rho_h v_h = \frac{N_h}{L^2}\frac{\sum_{j \in \{h\}} \underline{p}_j \cdot \underline{i}}{N_h m} = \frac{\sum_j \underline{p}_j^2/m}{L^2(2N-1)}\partial_{E_h}\sigma \tag{5.1}$$

where $j \in \{h\}$ means that $j$ is a species $h$ particle, and quantities of $O(N^{-1})$ have been neglected (see (2.3)) and $\sigma$ is $\frac{1}{k_B}$ times the entropy production rate, see (4.6),(4.7) and (4.8). Hence if we define $T(x)$, for each configuration $x$, by $\sum_j \frac{p_j^2}{2m} = NkT(x)$ and:

$$j_h \stackrel{def}{=} \frac{2N}{2N-1}\langle\frac{J_h}{kT}\rangle = \lim_{\tau\to\infty}\frac{1}{L^2}\frac{\sum_j \overline{\Lambda}_{u,\tau}^{-1}(x_j)\delta_\tau^{-1}(x_j)\,\partial_{E_h}\overline{\sigma}_\tau(x_j)}{\sum_j \overline{\Lambda}_{u,\tau}^{-1}(x_j)\delta_\tau^{-1}(x_j)} \tag{5.2}$$

where $\overline{\sigma}_\tau(x) = \frac{1}{\tau}\sum_{r=-\tau/2}^{\tau/2-1}\sigma(S^r x)$, see (4.7). If we recall (4.3) with $\frac{1}{2}\tau = M$ we see that $j_h$ can be regarded as the SRB average of $\frac{J_h}{kT}$, $h = 1, 2$.

This expression is similar to the formula derived from the generating function of the Helfand moments in [GD2], [GD1]: but it is not the same because in [GD2] the SRB is represented by using the notion of $(\varepsilon, \tau)$–separated sets (which are a somewhat more primitive or less concrete version of the parallelograms of the Markov partitions).

We shall also define $l_{u,\tau}, l_{s,\tau}$ as:

$$\overline{\Lambda}_{u,\tau}^{-1}(x)\delta_\tau(x)^{-1} = e^{\tau l_{u,\tau}(x)}, \qquad \overline{\Lambda}_{s,\tau}(x)\delta_{-\tau}^{-1}(x) = e^{\tau l_{s,\tau}(x)} \tag{5.3}$$

so that (4.5),(4.6) imply $l_{u\tau}(x) - l_{s\tau}(x) = \tau t_0 \overline{\sigma}_\tau(x)$. Hence we see that, if $\partial_k \equiv \partial_{E_k}$:

$$\begin{aligned}\partial_k j_h =& \frac{1}{L^2}\frac{\sum_j \overline{\Lambda}_{u,\tau}^{-1}(x_j)\delta_\tau^{-1}(x_j)\big(\partial_{hk}\overline{\sigma}_\tau(x_j) + \tau\partial_k l_{u,\tau}(x_j)\partial_h\overline{\sigma}_\tau(x_j)\big)}{\sum_j \overline{\Lambda}_{u,\tau}^{-1}(x_j)\delta_\tau^{-1}(x_j)}+\\ &-\frac{1}{L^2}\frac{\sum_j \overline{\Lambda}_{u,\tau}^{-1}(x_j)\delta_\tau^{-1}(x_j)\tau\partial_k l_{u\tau}(x_j)}{\sum_j \overline{\Lambda}_{u,\tau}^{-1}(x_j)\delta_\tau^{-1}(x_j)}\cdot\frac{\sum_j \overline{\Lambda}_{u,\tau}^{-1}(x_j)\delta_\tau^{-1}(x_j)\partial_h\overline{\sigma}_\tau(x_j)}{\sum_j \overline{\Lambda}_{u,\tau}^{-1}(x_j)\delta_\tau^{-1}(x_j)}\\ =&\frac{1}{L^2}\langle\partial_{hk}\overline{\sigma}_\tau\rangle - \frac{\tau}{L^2}\big(\langle\partial_k l_{u,\tau}\partial_h\overline{\sigma}_\tau\rangle + \langle\partial_k l_{u,\tau}\rangle\langle\partial_h\overline{\sigma}_\tau\rangle\big)\end{aligned} \tag{5.4}$$

Here we have interpreted the derivatives with respect to $E_h$ of $\overline{\sigma}_\tau(x)$ and $l_{u\tau}(x)$ by regarding $x$ as $\underline{E}$ independent. However this is *not* quite correct: in fact it is clear that we must consider such functions as defined on the attractor, not on the full phase space. The attractor depends on $\underline{E}$: it can in fact be identified with the unstable manifold $W_O^u$ of a fixed point $O$ or of a periodic orbit $O$ (see [GC2], §4): hence the point $x_j$, which has to be thought as a point on the attractor, will change with $\underline{E}$ even though it keeps the same symbolic representation (note that the Markov partition changes with $\underline{E}$ although the compatibility matrix does not, by the structural $\underline{E}$ stability theorem of Anosov, [AA], at least if $\underline{E}$ is small).

In taking the derivatives with respect to $\underline{E}$ of $l_{u\tau}(x_j)$ and in defining the current as $\partial_{E_h}\overline{\sigma}_\tau(x_j)$ there are therefore additional contributions proportional to $\partial_{E_h} x$. The latter quantity can be considered as a function of the symbolic history of $x$, *i.e.* as the function $\partial_{E_h} x(\underline{j})$ and in [G4] it is conjectured that:

*Conjecture: The function $\partial_{E_h} x(\underline{j})$ is a local function in the sense of the second theorem in §5 for all Anosov systems, or axiom A systems, depending smoothly on parameters $\underline{E}$.*

Assuming the validity of the conjecture and using it to perform rigorously an interchange of limits one can check, see [G4] for details, that the extra terms in the $\underline{E}$-derivatives of $\overline{\sigma}_\tau(x(\underline{j})), l_{u,\tau}(x(\underline{j}))$ at fixed history $\underline{j}$ just discussed *give no contribution* to the end result, *i.e.* they do not alter the validity of Onsager's reciprocity or of the fluctuation dissipation relation, derived below. Therefore, to avoid formal intricacies, we shall not take into account the extra terms and we proceed by ignoring them in (5.4) as well as in the following. The above conjecture has a mathematical nature and I do not discuss its proof here: I have not attempted to prove it (it seems closely related to Anosov's structural stability theorem, see [AA]).

By using the time reversal invariance we see that:

$$\langle \partial_k l_{u,\tau} \partial_h \overline{\sigma}_\tau \rangle = \frac{\sum \overline{\Lambda}_{u,\tau}^{-1} \delta_\tau^{-1} \partial_k l_{u,\tau} \partial_h \overline{\sigma}_\tau}{Z} = \tag{5.5}$$

$$= \frac{\sum_j \left( \overline{\Lambda}_{u,\tau}^{-1}(x_j) \delta_\tau^{-1}(x_j) \partial_k l_{u,\tau}(x_j) \partial_h \overline{\sigma}_\tau(x_j) + \overline{\Lambda}_{u,\tau}^{-1}(ix_j) \delta_\tau^{-1}(ix_j) \partial_k l_{u,\tau}(ix_j) \partial_h \overline{\sigma}_\tau(ix_j) \right)}{2Z}$$

where $Z$ denotes the "partition sum", *i.e.* the sum in the denominator of (5.4), and the averages are with respect to the distribution $\mu_{\tau/2,\tau}$.

Recalling that (see (4.5), (4.6)) $l_{u,\tau}(ix) = l_{s,\tau}(x)$, $\overline{\sigma}_\tau(ix) = -\overline{\sigma}_\tau(x)$ this becomes:

$$\frac{\sum_j \left( \overline{\Lambda}_{u,\tau}^{-1}(x_j) \delta_\tau^{-1}(x_j) \partial_k l_{u,\tau}(x_j) - \overline{\Lambda}_{s,\tau}^{-1}(x_j) \delta_{-\tau}^{-1}(x_j) \partial_k l_{s,\tau}(x_j) \right) \partial_h \overline{\sigma}_\tau(x_j)}{2Z} \tag{5.6}$$

The derivatives at $E_1 = E_2 = 0$ can be computed immediately by noting that *in such case*, $\overline{\Lambda}_{u,\tau}(x) \overline{\Lambda}_{s,\tau}(x) \frac{\delta_\tau(x)}{\delta_{-\tau}(x)} \equiv 1$ (see (4.6)). If we use that (4.6) implies $l_{u,\tau} - l_{s,\tau} = t_0 \tau \overline{\sigma}_\tau$ then it follows, from (5.6), that:

$$\left( \langle \partial_k l_{u,\tau} \partial_h \overline{\sigma}_\tau \rangle - \langle \partial_k l_{u,\tau} \rangle \langle \partial_h \overline{\sigma}_\tau \rangle \right) \Big|_{\underline{E} = \underline{0}} = \frac{\tau t_0}{2} \left( \langle \partial_k \overline{\sigma}_\tau \partial_h \overline{\sigma}_\tau \rangle - \langle \partial_k \overline{\sigma}_\tau \rangle \langle \partial_h \overline{\sigma}_\tau \rangle \right) \Big|_{\underline{E} = \underline{0}} \tag{5.7}$$

We also see that (since $\langle \partial_{hk} \overline{\sigma}_\tau \rangle$ vanishes in the present case):

$$\partial_k j_h = \lim_{\tau \to \infty} \frac{t_0}{2\tau L^2} \sum_{m=-\tau/2}^{\tau/2-1} \sum_{n=-\tau/2}^{\tau/2-1} \left( \langle \partial_k \sigma(S^m \cdot) \partial_h \sigma(S^m \cdot) \rangle - \langle \partial_k \sigma(S^m \cdot) \rangle \langle \partial_h \sigma(S^m \cdot) \rangle \right) \tag{5.8}$$

where the averages in the r.h.s. are with respect to $\mu_{\tau/2,\tau}$.

Hence we see that, apart from a further problem of interchange of limits (see below), (5.8) becomes:

$$\partial_k j_h = \frac{t_0}{2L^2} \sum_{m=-\infty}^{\infty} \left( \langle \partial_k \sigma(S^m \cdot) \partial_h \sigma(\cdot) \rangle - \langle \partial_k \sigma(\cdot) \rangle \langle \partial_h \sigma(\cdot) \rangle \right) \tag{5.9}$$

where the averages are with respect to the SRB distribution (*i.e.* to the limit of $\mu_{\tau/2,\tau}$).

The problem of interchange of limits is easily solved: under our assumption that the system is a transitive Anosov system the correlations of smooth observables decay exponentially (because they become local observables in the symbolic dynamics interpretation of the evolution, provided by the Markov partitions), not only for $\mu$ but also for $\mu_{\tau/2,\tau}$ (in the natural sense in which this may be interpreted in a finite $\tau$ case; *e.g.* by regarding the interval $[-\frac{\tau}{2}, \frac{\tau}{2}]$ as a circle), and uniformly in $\tau$.

The relation (5.9) implies that setting $L_{hk} = \langle \partial_h j_k \rangle |_{\underline{E} = \underline{0}}$ then:

$$L_{hk} = L_{kh} \tag{5.10}$$

follows. Note that (5.9) expresses the *fluctuation dissipation* relation between the transport matrix $L$ and the current–current equilibrium correlation.

In the case of the model in §4 the situation is similar. If $\sum_j \frac{p_j^2}{m} = 2N k_B T(x)$ and $J_q$ denotes the heat $-q_+ = -Q_+/L^2$ received by the gas from the thermostat $\mathcal{R}_+$ per unit time and volume:

$$\begin{aligned} \frac{J}{k_B T} &= \frac{2N-1)}{L^2} \frac{\sum_j \underline{p}_j \cdot \underline{i}/m}{\sum \underline{p}_j^2/m} = \frac{1}{L^2} \partial_E \sigma \\ \frac{1}{T_+} \frac{J_q}{T_+} &= \frac{2N_+ - 1}{L^2} \frac{-Q_+}{2N_+ k_B T_+^2} = \frac{1}{L^2} \partial_{\delta T} \sigma \end{aligned} \tag{5.11}$$

Hence the above argument yields, for the model in §4:

$$\partial_{\delta T} \langle \frac{J}{k_B T} \rangle \Big|_{\delta T, E=0} = \partial_E \langle \frac{1}{T} \frac{J_q}{k_B T} \rangle \Big|_{\delta T, E=0} \tag{5.12}$$

In general we can consider changing two parameters denoted $a, b$, *thermodynamic forces*, which control equations of motion of the system. Suppose that the entropy generation per unit time $\sigma$ has the form:

$$\sigma = \sum_r D_r \frac{Q_r}{\sum^r \underline{p}_j^2 / m} \tag{5.13}$$

where $\sum^r$ denotes that the coordinates $\underline{q}_j$ are coordinates of a particle belonging to a group of $N_r$ particles whose phase points are constrained by the $r$-th constraint that we impose on the system (to fix the coordinates that would evolve with a zero Lyapunov exponent, in the thermodynamic limit). And let $D_r$ be the number of degrees of freedom of the $r$-th group of particles (in 2 space dimensions $D_r \simeq 2N_r$); then the above argument can be immediately generalized to yield that the *flows* $J_a = \langle \partial_a \sigma \rangle$ and $J_b = \langle \partial_b \sigma \rangle$ verify:

$$\partial_b J_a \Big|_{a,b=0} \stackrel{def}{=} L_{12} = L_{21} \stackrel{def}{=} \partial_a J_b \Big|_{a,b=0} \tag{5.14}$$

which is a general Onsager reciprocity relation between "thermodynamic forces" and "currents". From (5.7) we also see that the matrix $\partial_b J_a$ is positive definite.

Note that, as mentioned above, in defining $\partial_a \sigma, \partial_b \sigma$ one has really to think of $\sigma$ as defined on the space of the symbolic sequences $\sigma = \sigma(x(\underline{j}))$ (so that $\partial_a \sigma = \frac{\partial \sigma}{\partial a}(x(\underline{j})) + \frac{\partial \sigma}{\partial x} \frac{\partial x(\underline{j})}{\partial a}$: this conceptually different from the "naive" $\frac{\partial \sigma}{\partial a}(x(\underline{j}))$ although (*in the above models* it does not affect the end result).

### §6. Remarks.

(1) The models in §3,§4 have been considered as undergoing transformations at constant energy $U$. This is not very satisfactory as one also, and mainly, wants to understand also transformations in which the internal energy is allowed to change. According to the ideas in §2 this case can be treated by imposing that $U$ is constant; but the constant value is fixed on the basis of the *equation of state* of the system. The latter is the relation linking $U$ to the other system parameters:

$$U = f(E^1, E^2), \qquad U = f(E, T_-, T_+) \tag{6.1}$$

in the cases of the models of §3,§4. Here $f$ is determined by the dynamics itself, but its computation requires mathematical difficulties that we cannot expect to be able to solve (in general).

Nevertheless we can proceed as in the previous section: in taking the derivatives with respect to the parameters there will be extra terms that arise from the fact that the energy surface changes as described by (6.1), but the basic symmetry $L_{hk} = L_{kh}$ remains, as one can check (by taking also into account that in the equilibrium state obtained when the thermodynamic forces, *i.e.* the variations from the equilibrium values of the parameters, are set to 0 then the currents vanish).

(2) It is quite clear that the discussion of the previous sections can be extended to many other models. But it is not clear how far one can really extend the considerations. For instance it would be desirable to extend, if possible, the considerations to a microscopic model of a macroscopic continuum obeying the macroscopic equations for a fluid or a mixture of fluids (possibly with chemical reactions allowed), as defined in [DGM].

(3) Onsager relations are often regarded to be consequences solely of the reversibility of the equilibrium dynamics, (see, however, [DJL]). Therefore they *must* hold also for our models simply because they could be derived "as usual", see [DGM], p. 100-101, for a "usual derivation".

Hence it is worth pointing out that the "usual derivation" rests on several assumptions, none of which is needed *if* the chaotic hypothesis is retained, at least in a dynamics of the type considered in the above models.

(4) The "usual derivation" assumptions are the following;

(a) *linear regression*, *i.e.* the "time behaviour of the state parameters can be described by linear equations", see [DGM], p. 36 and p.100.
(b) The "Boltzmann postulate": *i.e.* the equilibrium distribution is such that the "state parameters" $\underline{a}$ obey a gaussian *large deviation law*, see [DGM], p. 91 eq. (46). This means that they have an equilibrium probability distribution $p(\underline{a})$ proportional to $e^{-\frac{1}{2k_B} G \underline{a} \cdot \underline{a}} = e^{S(\underline{a})/k_B}$ where $S$ is the "entropy" of the state with state parameters $\underline{a}$. The entropy is *defined* by the above formula but it is treated as if it had the properties we may expect from a genuine equilibrium entropy function (note that this is an assumption). By "large" one means an amount much larger than the root mean square values at equilibrium (see [DGM], p. 100), *i.e.* macroscopic (but still very small). One should remark that the gaussian nature *is not* a consequence of a normal distribution assumption as the latter usually concerns *small deviation* of the order of the root mean square values. On the other hand a careful examination of the proof shows that all is really needed is that $-S(\underline{a})$ is a convex smooth function near $\underline{a} = \underline{0}$. Therefore the really strong part of the assumption above is that $S(\underline{a})$ *behaves as an ordinary entropy function* (a concept that would require some more precise definition).
(c) The equilibrium evolution is *reversible*.

Then it follows that the time evolution of a fluctuation is such that the "state parameters" $\underline{a}$ verify $\underline{\dot{a}} = L \underline{X}$ with $\underline{X} = -G \underline{a}$ and the symmetry $L = L^T$, see [DGM], p. 101÷102.

An initial (distribution of) microscopic configurations, close to the equilibrium state, generates a macroscopic state in which fluctuations are possible: so one can consider the free evolution of a state in which the initial "state parameters" have an average value off by $\underline{a}$ from the equilibrium value $\underline{0}$. The evolution will then verify the above "symmetry" relation.

For instance $\underline{a} = (a_1, a_2)$ could be the horizontal center of mass coordinates of the two species of particles in model 1 above.

(5) The connection with "reality" requires further assumptions. Considering our model 1 for definiteness, suppose that we act on the system with small forces thus driving an evolution of the average values of the "state parameters" $t \to \underline{a}(t)$, and creating an entropy per unit time $\frac{(\dot{a}_1 N_1 E^1 + \dot{a}_2 N_2 E^2)}{T}$ (see §2: this makes sense for small fields when the temperature $T$ can be identified with the average kinetic energy per particle). Note also that, by (b) above, the entropy creation rate is in general $\dot{S} = -G \underline{a} \cdot \underline{\dot{a}} \equiv \underline{X} \cdot \underline{\dot{a}}$ with $\underline{X} = -G\underline{a}$.
(d) Then Onsager supposes (see [O]: "As before we shall assume that the average regression of fluctuations obey the same laws as the corresponding macroscopic irreversible processes") that a fluctuation forced by external forces evolves as if it had occurred spontaneouly; *i.e.* if $\underline{X}$ is given, $\underline{X} = \frac{(N_1 E^1, N_2 E^2)}{T}$, recalling that $N_1 = N_2 = \frac{N}{2}$, it is $\underline{\dot{a}} = \frac{N}{2} L \frac{\underline{E}}{T}$ (or $\underline{j} = \frac{N}{2} L \frac{\underline{E}}{T}$, in the notations of the present paper), with $L_{12} = L_{21}$. Note that this is done (and can only be correct) up to corrections $O(\underline{E}^2)$.

Other derivations are more fundamental and are based on kinetic theory (see [C], [DGM]) or on the pure microscopic dynamics ([DGM], [D]), but still they rely on various assumptions besides time reversibility of the equilibrium dynamics.

(6) With our chaotic hypothesis all the above assumptions (a)÷(d) are *not necessary*, if one considers the models in §2,§3, because the final result (*i.e.* $\underline{j}$ proportional to $L\,\underline{E}$ with $L$ symmetric and positive definite) has been drawn without further hypotheses (other than the mathematical conjecture in §5). Of course assuming the reversibility of the dynamics *even* in nonequilibrium (close to equilibrium) *and* the chaotic hypothesis is in some sense much stronger than the assumptions (a)÷(d) (but note that the reversibility in nonequilibrium , close to equilibrium, is a form of the assumption (d)).

It has, *however*, the basic advantage of being a conceptually simple general assumption for nonequilibrium statistical mechanics, which should furthermore be valid without even the restriction of being close to equilibrium.

As a final remark one should add that, although reversibility of the nonequilibrium dynamics is assumed in this paper, it is quite likely that the ideas and methods can be extended to genuinely non reversible models. Attempts at such applications can be already found in [GC1],[GC2],[G4] and I hope that they can be generalized to more general physical situations quite beyond the, so far special, cases mentioned.

A much debated question is the extension of the Onsager relations to the really nonequilibrium regime (*i.e.* not close to equilibrium). In this case the very notion of reciprocity becomes ill defined: recently the proposal of interpreting the relations as the statement that the flow of the state parameters is, in a suitable sense, a "gradient flow" has emerged, [GJL]. The chaotic hypothesis in principle applies also in the nonlinear regimes but it is unlikely that it can be as powerful a tool to cover deterministic versions of the concrete, exactly treated, stochastic models of [GJL].

**Appendix A1:** *The Gauss minimal constraint principle.*

Let $\varphi(\underline{\dot{x}},\underline{x}) = 0$ be a constraint and let $\underline{R}(\underline{\dot{x}},\underline{x})$ be the constraint reaction and $\underline{F}(\underline{\dot{x}},\underline{x})$ the active force.

Consider all the possible accelerations $\underline{a}$ compatible with the constraints and a given initial state $\underline{\dot{x}},\underline{x}$. Then $\underline{R}$ is *ideal* or *verifies the principle of minimal constraint* if the actual accelerations $\underline{a}_i = \frac{1}{m_i}(\underline{F}_i + \underline{R}_i)$ minimize the *effort*:

$$\sum_{i=1}^{N} \frac{1}{m_i}(\underline{F}_i - m_i\,\underline{a}_i)^2 \longleftrightarrow \sum_{i=1}^{N}(\underline{F}_i - m_i\,\underline{a}_i)\cdot\delta\underline{a}_i = 0 \qquad (A1.1)$$

for all possible variations $\delta\underline{a}_i$ compatible with the constraint $\varphi$.

Since all possible accelerations following $\underline{\dot{x}},\underline{x}$ are such that $\sum_{i=1}^{N} \partial_{\underline{\dot{x}}_i}\varphi(\underline{\dot{x}},\underline{x})\cdot\delta\underline{a}_i = 0$ we can write:

$$\underline{F}_i - m_i\,\underline{a}_i - \alpha\,\partial_{\underline{\dot{x}}_i}\varphi(\underline{\dot{x}},\underline{x}) = \underline{0} \qquad (A1.2)$$

with $\alpha$ such that $\frac{d}{dt}\varphi(\underline{\dot{x}},\underline{x}) = 0$, *i.e.* :

$$\alpha = \frac{\sum_i (\underline{\dot{x}}_i\cdot\partial_{\underline{\dot{x}}_i}\varphi + \frac{1}{m_i}\underline{F}_i\cdot\partial_{\underline{\dot{x}}_i}\varphi)}{\sum_i m_i^{-1}(\partial_{\underline{\dot{x}}_i}\varphi)^2} \qquad (A1.3)$$

which is the analytical expression of the Gauss' principle.

*Acknowledgements:* I am indebted to J.L. Lebowitz, G.L. Eyink and G. Jona for very helpful comments and in particular to E.G.D. Cohen for many very valuable and inspiring suggestions and hints and for his constant interest and encouragement. This work is part of the research

References.

[AA] Arnold, V., Avez, A.: *Ergodic problems of classical mechanics*, Benjamin, 1966.

[Bo] Bowen, R.: *Equilibrium states and the ergodic theory of Anosov diffeomorphisms*, Lecture notes in mathematics, vol. **470**, Springer Verlag, 1975.

[BEC] Baranyai, A., Evans, D.T., Cohen, E.G.D.: *Field dependent conductivity and diffusion in a two dimensional Lorentz gas*, Journal of Statistical Physics, **70**, 1085–1098, 1993.

[C] Cohen, E.G.D.: , H.J.M. Hanley, ed., Marcel Dekker, 1969.

[CG1] Gallavotti, G., Cohen, E.G.D.: *Dynamical ensembles in nonequilibrium statistical mechanics*, Physical Review Letters, **74**, 2694–2697, 1995.

[CG2] Gallavotti, G., Cohen, E.G.D.: *Dynamical ensembles in stationary states*, in print in Journal of Statistical Physics, 1995.

[D] Dorfman, J.R.: *Transport coefficients*, entry for the "Enciclopedia delle Scienze Fisiche", Enciclopedia Italiana, Roma, 1995 (in italian, preprint in english available from the author), p. 1-31.

[DGM] de Groot, S., Mazur, P.: *Non equilibrium thermodynamics*, Dover, 1984.

[E] Eyink, G.: *Turbulence noise*, in *mp_arc@ math. utexas. edu*, #95-254, 1995.

[ECM1] Evans, D.J., Cohen, E.G.D., Morriss, G.P.: *Viscosity of a simple fluid from its maximal Lyapunov exponents*, Physical Review, **42A**, 5990–5997, 1990.

[ECM2] Evans, D.J., Cohen, E.G.D., Morriss, G.P.: *Probability of second law violations in shearing steady flows*, Physical Review Letters, **71**, 2401–2404, 1993.

[ELS] Eyink, G.L., Lebowitz, J.L., Spohn, H.: *Hydrodynamics and Fluctuations Outside of Local Equilibrium: Driven Diffusive Systems*, in *mp_arc@math.utexas.edu*, #95-168.

[EM] Evans, D.J., Morriss, G.P.: *Statistical Mechanics of Nonequilibrium fluids*, Academic Press, New York, 1990.

[EMY] Esposito, R., Marra, R., Yau, H.T.: *Diffusive Limit of Asymmetric Simple Exclusion*, In: "The State of Matter", ed. M. Aizenmann, H. Araki., World Scientific, Singapore, 1994.

[ES] Evans, D.J., Searles, D.J.: *Equilibrium microstates which generate second law violating steady states*, Research school of chemistry, Canberra, ACT, 0200, preprint, 1993.

[G1] Gallavotti, G.: *Topics in chaotic dynamics*, Lectures at the Granada school, ed. Garrido–Marro, Lecture Notes in Physics, **448**, 1995.

[G2] Gallavotti, G.: *Reversible Anosov diffeomorphisms and large deviations.*, Mathematical Physics Electronic Journal, **1**, 1-12, 1995.

[G3] Gallavotti, G.: *Coarse graining and chaos*, in preparation.

[G4] Gallavotti, G.: *Chaotic principle: some applications to developed turbulence*, in *mp_arc @math. utexas. edu*, #95-232, 1995.

[GCDR] Garcia–Colin, L.S., Del Rio, J.L.: *Green's contributions to nonequilibrium statistical mechanics*, in "Perspectives in statistical physics: M.S. Green memorial volume", ed. H.J. Ravechè, p. 75–87, North Holland, 1981.

[GD1] Gaspard, P., Dorfman, J.R.: *Chaotic scattering theory, of transport and reaction rate ccoefficients*, Physical Review, **E51**, 28–35, 1995.

[GD2] Gaspard, P., Dorfman, J.R.: *Chaotic scattering theory, thermodynamic formalism and transport coefficients*, in *chao-dyn@xyz.lanl.gov*, # 9504014.

[GJL] Gabrielli, D., Jona–Lasino, G., Landim, C.: *Microscopic reversibility and thermodynamic fluctuations*, in *mp_arc@ math. utexas. edu*, #95-248, 1995.

[HHP] Holian, B.L., Hoover, W.G., Posch. H.A.: *Resolution of Loschmidt's paradox: the origin of irreversible behavior in reversible atomistic dynamics*, Physical Review Letters, **59**, 10–13, 1987.

[KV] Kipnis, C., Olla, S., Varadhan, S.R.S., , Communications in Pure and Applied Mathematics, **XLII**, 243–, 1989.

[LA] Levi-Civita, T., Amaldi, U.: *Lezioni di Meccanica Razionale*, Zanichelli, Bologna, 1927 (reprinted 1974), volumes $I, II_1, II_2$.

[LOY] Landim, C., Olla, S., Yau, H.T.; *First-order correction for the hydrodynamic limit of asymmetric simple exclusion processes in dimension $d \geq 3$*, preprint, Ecole Polytechnique, R. I. No. 307, Nov. 1994.

[LPR] Livi, R., Politi, A., Ruffo, S.: *Distribution of characteristic exponents in the thermodynamic limit*, Journal of Physics, **19A**, 2033–2040, 1986.

[O] Onsager, L: *Reciprocal relations in irreversible processes. II*, Physical Review, **38**, 2265–2279, 1932.

[PH] Posch, H.A., Hoover, W.G.: *Non equilibrium molecular dynamics of a classical fluid*, in "Molecular Liquids: New Perspectives in Physics and Chemistry", ed. J. Teixeira-Dias, Kluwer Academic Publishers, p. 527–547, 1992.

[R1] Ruelle, D.: *Chaotic motions and strange attractors*, Lezioni Lincee, notes by S. Isola, Accademia Nazionale dei Lincei, Cambridge University Press, 1989; see also: Ruelle, D.: *Measures describing a turbulent flow*, Annals of the New York Academy of Sciences, **357**, 1–9, 1980. For more technical expositions see Ruelle, D.: *Ergodic theory of differentiable dynamical systems*, Publications Mathémathiques de l' IHES, **50**, 275–306, 1980.

[R2] Ruelle, D.: *A measure associated with axiom A attractors*, American Journal of Mathematics, **98**, 619–654, 1976.

[S] Sinai, Y.G.: *Gibbs measures in ergodic theory*, Russian Mathematical Surveys, **27**, 21–69, 1972. Also: *Lectures in ergodic theory*, Lecture notes in Mathematics, Princeton U. Press, Princeton, 1977.

[UF] Uhlenbeck, G.E., Ford, G.W.: *Lectures in Statistical Mechanics*, American Mathematical society, Providence, R.I., pp. 5,16,30, 1963.